\documentclass[preprint,prd,aps,superscriptaddress,11pt]{revtex4-1}

\usepackage{graphicx}
\usepackage{amsmath}
\usepackage{amssymb}
\usepackage{dcolumn}
\usepackage{bm}
\usepackage{color}
\usepackage{float}

\newcommand{\be}{\begin{equation}}
\newcommand{\ee}{\end{equation}}
\newcommand{\ba}{\begin{eqnarray}}
\newcommand{\ea}{\end{eqnarray}}
\newcommand{\no}{\nonumber \\}
\newcommand{\gsim}{\mathrel{\hbox{\rlap{\lower.55ex \hbox {$\sim$}}
                   \kern-.3em \raise.4ex \hbox{$>$}}}}
\newcommand{\lsim}{\mathrel{\hbox{\rlap{\lower.55ex \hbox {$\sim$}}
                   \kern-.3em \raise.4ex \hbox{$<$}}}}

\def\roughly#1{\mathrel{\raise.3ex\hbox{$#1$\kern-.75em%
\lower1ex\hbox{$\sim$}}}}
\def\lsim{\roughly<}
\def\gsim{\roughly>}

\def\vx{{\vec x}}

\def\({\left(}
\def\){\right)}
\def\[{\left[}
\def\]{\right]}
\def\<{\langle}
\def\>{\rangle}

\def\k{{\kappa}}
\def\l{{\lambda}}

\def\d{{\delta}}

\def\o{{\omega}}

\def\e{{\epsilon}}

\def\b{{\beta}}
\def\c{{\chi}}

\def\g{{\gamma}}

\def\m{{\mu}}
\def\n{{\nu}}
\def\r{{\rho}}
\def\s{{\sigma}}

\def\t{{\tau}}

\def\0{{(0)}}
\def\1{{(1)}}
\def\2{{(2)}}

\newcommand{\pd}{{\partial}}
\newcommand{\pr}{\parallel}

\begin{document}

\title{\bf Thermal Diffusion and Quantum Chaos in Neutral Magnetized Plasma}

\author{Wei Li}
\email{r07222072@ntu.edu.tw}
\affiliation{Department of Physics and Astronomy, National Taiwan University, Taipei 10617, Taiwan}
\author{Shu Lin}
\email{linshu8@mail.sysu.edu.cn}
\affiliation{School of Physics and Astronomy, Sun Yat-Sen University, Zhuhai 519082, China}
\author{Jiajie Mei}
\email{jiajiemei@outlook.com}
\affiliation{School of Physics and Astronomy, Sun Yat-Sen University, Zhuhai 519082, China}

\date{\today}

\begin{abstract}
We calculate the thermal diffusion constant $D_T$ and butterfly velocity $v_B$ in neutral magnetized plasma using holographic magnetic brane background. We find the thermal diffusion constant satisfies Blake's bound. The constant in the bound $D_T2\pi T/v_B^2$ is a decreasing function of magnetic field. It approaches one half in the large magnetic field limit. We also find the existence of a special point defined by Lyapunov exponent and butterfly velocity on which pole-skipping phenomenon occurs.
\end{abstract}

\maketitle


\newpage

\section{Introduction}

Recently, there have been remarkable progress in understanding the connection between transport properties and chaos in strongly coupled many-body systems. On one hand, the transport of many-body systems is generically governed by a relaxation time scale $\t$. While the mechanism of relaxation is diverse in different systems, the relaxation time scale is expected to be bounded by smallest time scale allowed by uncertainly principle $\t\gtrsim\hbar/k_BT$ \cite{1405.3651}. On the other hand, a similar bound exists for the scrambling rate of the system $\l_L\leq 2\pi k_BT/\hbar$, with $\l_L$ being the Lyapunov exponent defined through out-of-time-order correlator (OTOC) for two Hermitian operators $W$ and $V$ \cite{1306.0622,1409.8180,1412.6087}
\begin{align}\label{otoc}
  C(t,\vx)=-\<[W(t,\vx),V(0,0)]^2\>_\b\sim e^{\l_L(t-t_*-\frac{|\vx|}{v_B})}.
\end{align}
Here $v_B$ is the butterfly velocity and $t_*$ is the scrambling time. The bound on scrambling rate is conjectured to be $2\pi T$ \cite{1503.01409}, and the bound is saturated in holographic systems \cite{1306.0622} and SYK model \cite{1605.06098}. 

Interestingly the connections between transport and chaos can be made further. It has been proposed that diffusion constant, as a proxy of relaxation satisfies a stronger universal bound: $D\sim \t\sim \hbar v_B^2/k_BT$ \cite{1603.08510}. There have been evidence for this bound for charge diffusion and thermal diffusion in holographic systems \cite{1604.01754,1610.02669,1611.09380,1702.08803,1704.00947,1705.07896,1705.01766,1707.02843,1708.05691,1708.08822,1712.00705,1805.05351}, SYK models \cite{1609.07832,1702.08462,1705.09818,1711.07903,1904.02174}. However the bound for charge diffusion can be violated with derivative correction in holographic system \cite{1612.05500} and field theory models \cite{1611.00003,1612.00849}. The bound for thermal diffusion in general holds, see also exceptions \cite{1608.03286,1710.07896,1710.03738}. The non-universality/universality of the bounds can be qualitatively understood in holographic systems: the butterfly velocity is determined by shock wave solution, which only probes near horizon region of the gravity background. To determine the diffusion constant, we need to use Einstein relation. For charge diffusion, $D_c=\s/\c$. The conductivity $\s$ is also fixed by the Infra-red (IR) fixed point. However the susceptibility $\c$ is non-universal, but depends on the full bulk metric. Thus the bound for charge diffusion can be violated. The situation is different for thermal diffusion, $D_e=\k/c_\r$. Here thermal conductivity $\k$ is again fixed by IR fixed point. The specific heat $c_\r$ and $v_B$ are determined by deformation of the fixed point. It has been shown that the same irrelevant deformation enters $c_\r$ and $v_B$ \cite{1611.09380}.

In fact, \cite{1611.09380} represents one class of examples where the IR fixed point is $AdS_2\times R^2$. In this paper, we present another class where the IR fixed point is given by $BTZ\times R^2$. The corresponding holographic model is the magnetic brane solution \cite{0908.3875}, which interpolates between the IR fixed point to asymptotic $AdS_5$ in the ultra-violet (UV). The dual theory is neutral plasma subject to external magnetic field. The presence of external magnetic field introduces momentum dissipation to the systems through Lorentz force, turning the energy density from a propagating mode into a diffusive mode. This is exactly the thermal diffusion we use to test the bound. The validity of the bound for charge diffusion has been studied in \cite{1805.05351}.

Recently an intriguing phenomenon of pole-skipping in the correlator of energy density has been studied \cite{1710.00921,1801.00010,1809.01169,1904.12883,1904.12862}. At a special point $\o=i\l_L$ and $k=\frac{i\l_L}{v_B}$ set by chaotic quantities, one component Einstein equation is trivially satisfied. This gives rise to an extra independent solution to metric perturbation, leading to indeterminacy of the energy density correlator at the special point. Moreover, there is a line of poles of the correlator crossing the special point. The pole is skipped at precisely the special point. The existence of the special point and pole-skipping phenomenon shed new light on the connection between quantum chaos and transports. Interestingly, we also confirm the existence of the special point in the magnetic brane background. This provides another non-trivial evidence for the connection.

The paper is organized as follows: In Section II, we first present a magnetohydrodynamic derivation of the thermal diffusion mode. Then we calculate the thermal diffusion constant independently from the magnetic brane background. The results of the two approaches are in agreement with each other. Section III is devoted to the calculation of butterfly velocity from shock wave in the magnetic brane background. In Section VI, we test the bound on thermal diffusion constant based on results obtained in the preceding sections. The results are found to be analogous to the case of incoherent metal. In Section V, we show the existence of special point in magnetic brane background and discuss its implications for magnetohydrodynamics.
We conclude and discuss possible extensions in Section VI.

\section{Thermal diffusion in neutral magnetized plasma}

\subsection{Thermal Diffusion from Magnetohydrodynamics}

We first present the derivation of thermal diffusion constant in the framework of Magnetohydrodynamics (MHD) \cite{1703.08757,1610.07392} first. For simplicity we consider neutral plasma, in which the fluctuation of energy density and charge density decouple. We are interested in the diffusive mode of the energy density. It arises from fluctuation of fluid velocity perpendicular to magnetic field and energy density (or equivalently temperature). 
The constitutive equation can be expressed as follows:
\begin{align}\label{const_rel}
&j_i=\(E_i+\e _{ijk}v_jB_k\)\s _{\perp}+\partial _t P_i+\e _{ijk}\partial _j M_k, \no
&T^{00}=\e, \no
&T^{ii}=p_{\perp}=p_{\pr}-M B, \no
&T^{0i}=(\e +p_{\pr})v_i-\e_{ijk}E_jM_k.
\end{align}
Here the three terms in current density correspond to conducting current, polarization current and magnetization current. We consider plasma subject to constant external magnetic field and vanishing electric field. It follows that the polarization is induced by motion of fluid:
\begin{align}\label{polarization}
P_i=\e _{ijk}v_{j}M_k
\end{align}
The index $i$ in stress tensor components in \eqref{const_rel} is transverse to magnetic field.
Without loss of generality, we choose magnetic field to point in $z$ direction and fluctuating fluid velocity along $x$ direction. Assuming the plane wave form $e^{-i\omega t+ikx}$ for the fluctuations, 
%
we can write down \eqref{const_rel} explicitly:
\begin{align}\label{const_exp}
&j_y=-\s _{\perp}B v_x+i\omega M v_x-ik\frac{\pd M}{\pd T}\delta T, \no
&\delta T^{00}=(\frac{\pd \e}{\pd T})_T\delta T=c_v \delta T, \no
&\delta T^{xx}=\delta p-\delta M B=s\delta T-\frac{\pd M}{\pd T}B\delta T, \no
&T^{0x}=(\e +p_{\pr})v_x. 
\end{align}
The conservation equations are given by
\begin{align}
  &\pd_\m T^{\m\n}=j_\l F^{\l\n}, \no
  &\pd_\m j^\m=0,
 \end{align}
with the following nonvanishing components in our case
\begin{align}\label{conservation}
&-i\omega c_v\delta T +ik(\e +p)v_x=0, \no
&-i\omega (\e+p)v_x+iks\delta T-i\omega M B v_x+\s _{\perp}B^2v_x=0.
\end{align}
\eqref{const_exp} and \eqref{conservation} are just algebraic equations, which admit two solutions as dispersion relation. One is diffusive mode:
\begin{align}
\omega &=-\frac{i(\e +p)^2}{B^2 c_v T\s _{\perp}}k^2 \no
&=-\frac{i(\e +p)^2 }{B^2T^2\frac{\pd s}{\pd T}\s_{\perp}}k^2.
\end{align}
The other one is a gapped mode lying beyond the regime of MHD. 
\begin{align}
\omega=-\frac{i B^2\s _{\perp}}{p+\e+M B}+\frac{i(\e ^2+p^2+2\e p)k^2}{B^2 c_v T\s _{\perp}}
\end{align}
Thus we obtain the thermal diffusion constant as follows:
\begin{align}\label{D_mhd}
 D_T=\frac{(\e +p)^2 }{B^2T^2\frac{\pd s}{\pd T}\s_{\perp}}.
\end{align}
In general, MHD is expected to be valid in the regime $B\ll T^2$. Nevertheless we will confirm that this result holds at finite $B/T^2$ as well with explicit holographic model calculation.

\subsection{Thermal Diffusion from Holographic Calculation}

In this section, we present a holographic calculation of thermal diffusion constant in neutral magnetic brane background \cite{0908.3875}. This serves as a confirmation of the result obtained in the previous section based on magnetohydrodynamics. The neutral magnetic brane background is a solution of the Einstein-Maxwell action
\begin{align}
  S=\int d^5x\sqrt{-g}\(R+12-\frac{1}{4}F_{MN}F^{MN}\).
\end{align}
We have set the overall factor of the action to $1$ for simplicity as it does not affect the diffusion constant. We also set AdS radius to $1$.
The solution is given as follows
\begin{align}
  &ds^2=-Udt^2+\frac{dr^2}{U}+V_x\((dx^1)^2+(dx^2)^2\)+V_ydy^2,\no
  &F=B dx^1\wedge dx^2.
\end{align}
Here $U$, $V_x$ and $V_y$ are functions of radial coordinate $r$ only. The constant magnetic field $B$ is along $y$ direction. The background satisfies Einstein equation with backreaction of the gauge field $F$, which leads to anisotropic metric components. The background is dual to strongly coupled neutral plasma subject to magnetic field.
For plasma at temperature $T$, the background is a black hole with the warping function $U$ adopting the following expansion near horizon
\begin{align}
  U=4\pi T(r-r_h)+O\((r-r_h)^2\).
\end{align}
The solution needs to be obtained by numerically integrating the Einstein equation from horizon to boundary. A notable feature of this background is that it interpolates between IR fixed point $BTZ\times R^2$ at the horizon to UV fixed point $AdS_5$ at the boundary.

In general, finding transport coefficients like diffusion constant amounts to solving linearized bulk equations, which could be complicated for magnetic brane background. However, it has been showed by Donos, Gauntlett and Ziogas (DGZ) that for a general class of background, the procedure can be simplified as solving constraint equations near the horizon. The idea of DGZ is to start with thermodynamic normal mode, which consists of constant perturbations in field theory directions. The normal mode is then promoted to hydrodynamic mode by including small momenta and frequency. The constraint equations are sufficient to fix the dispersion relation order by order in small momenta expansion.
We will closely follow this method. For notational simplicity, we shift the horizon to $r=0$ below.
To proceed, we need to turn on linear perturbation for metric and gauge field $g_{MN}\to g_{MN}+h_{MN}$, $A_M\to A_M+a_M$. Motivated by the MHD analysis, we turn on only a subset of the perturbations. The metric and gauge field perturbations are chosen with the following gauge condition and satisfy the following expansion near horizon:
\begin{align}\label{pert}
&h_{tt}=e^{-i\omega v}(4\pi Tr)\(h_{tt}^\0(x)+O(r)\), \no
&h_{rr}=e^{-i\omega v}\frac{1}{4\pi Tr}\(h_{rr}^\0(x)+O(r)\), \no
&h_{ij}=e^{-i\omega v}\(h_{ij}^\0(x)+O(r)\), \no
&h_{tr}=e^{-i\omega v}\(h_{tr}^\0(x)+O(r)\), \no
&h_{ti}=e^{-i\omega v}\(h_{ti}^\0(x)+r h_{ti}^\1(x)+O(r^2)\),  \no
&h_{ri}=e^{-i\omega v}\frac{1}{4\pi Tr}\(h_{ri}^\0(x)+rh_{ri}^\1(x)+O(r^2)\), \no
&a_i=e^{-i\omega v}\(a_i^\0(x)+O(r)\),
\end{align}
with
\begin{align}\label{gc}
  &2\pi T\(h_{tt}^\0(x)+h_{rr}^0(x)\)=-4\pi T h_{rt}^\0(x)=p(x),\no
  &h_{ti}^\0(x)=h_{ri}^\0(x)=-v_i(x).
\end{align}
Where the $ij$ represents the spatial component and $v$ is the ingoing Eddington-Finkelstein coordinate $v=t+\frac{ln r}{4\pi T}$. The argument $x$ of the functions with superscript $\0$ refers to spatial coordinates only.

The master equations are the constraint equations in the Hamiltonian formulation of holographic models \cite{1106.4826}. Recall the ADM decomposition of metric
\begin{align}\label{ADM}
  ds^2=N^2dr^2+\g_{\m\n}(dx^\m+N^\m dr)(dx^\n+N^\n dr),
\end{align}
with $\m=(t,x^1,x^2,y)$.
In the presence of perturbations, the functions in the ADM form are given by
\begin{align}
  N=U^{-1/2}\(1+\frac{1}{2}U h_{rr}\),\quad N_t=-\frac{1}{U}h_{rt},\quad N_i=h_{ri}.
\end{align}
In Hamiltonian formulation, we define stress tensor and charge current on each $r$-hypersurface as
\begin{align}\label{def_KJ}
&\pi ^{\mu \nu}=K h^{\mu \nu}-K^{\mu \nu}, \no
&J^{\mu}=-N^{-1}(F_{r}^\m-N_\n F^{\n\m}).
\end{align}
Here we define these without $\sqrt{-\gamma}$ as tensor and vector with respect to the induced metric $\g_{\m\n}$. The extrinsic curvature is defined as $K_{\m\n}=\frac{1}{2}\(\dot{\g}_{\m\n}-\nabla_\m N_\n-\nabla_\n N_\m\)$.
With these definitions, the constraint equations are energy-momentum conservation and charge conservation, which can be written as
\begin{align}\label{constraint_general}
  &H_{\mu}\equiv 2\nabla _{\nu}\pi^{\nu}_{\mu}-F_{\mu \nu} J^{\nu}=0, \no
  &C=\nabla_\m J^\m=0.
\end{align}
Plugging the perturbations \eqref{pert} into \eqref{constraint_general} and expanding terms to leading order in $r$, we end up with the following equations
\begin{align}\label{explicit}
  &\pd_iQ_\0^i-i\o(2\pi T) \sqrt{-g_\0}g^{ij}_\0h_{ij}^\0=0, \no
  &-2\pd^j\pd_{[j}v_{i]}+\(1+\frac{i\o}{4\pi T}\)\pd_ip+i\o\(-h_{ti}^\1+\pd_ih_{tr}^\0-g_{ik}^\1v^k+\pd^kh_{ki}^\0+\frac{i\o}{4\pi T}\(h_{ti}^\1-h_{ri}^\1\)\) \no
  &+F^{ik}_\0\(-g^{lj}_\0F_{jk}^\0v_l-i\o a_k^\0\)=0, \no
  &i\o\pd_i\(g^{ij}_\0a_j^\0\)=0,
\end{align}
with $Q_\0^i=4\pi T\sqrt{-g_\0}v^i$ being the thermal current. The three equations correspond to energy conservation, momentum conservation and charge conservation respectively.

Let us begin with the thermodynamic normal mode. This is obtained by an infinitesimally constant shift of temperature $\d T$ of the background solution. In neutral background we consider here, the thermal current decouples from charge current, so that we do not need to consider shift in chemical potential for the normal mode. The temperature shift is by construction a normal mode. The corresponding nonvanishing perturbations are given by
\begin{align}\label{TH}
  h_{tt}^{TH}=-4\pi r\d T,\quad h_{rr}^{TH}=-\frac{\d T}{4\pi T^2r},\quad h_{ij}^{TH}=\frac{\pd g_{ij}^\0}{\pd T}\d T.
\end{align}
However, this normal mode does not satisfy the gauge condition \eqref{gc}. We can achieve the gauge condition by a coordinate transformation $t\to t+\frac{\d T}{T}g(r)$. With properly chosen function $g(r)=\ln r/(4\pi T)+g^\1 r$ near the horizon, the normal mode transformed to the following form near the horizon
\begin{align}\label{RT}
  h_{tt}^{RT}=-4\pi r\d T,\quad h_{rr}^{RT}=-\frac{\d T}{4\pi T^2r},\quad h_{tr}^{RT}=-\frac{\d T}{T}, \quad h_{ij}=\frac{\pd g_{ij}^\0}{\pd T}\d T.
\end{align}
We can read off the coefficients $h_{tt}^\0=h_{rr}^\0=h_{tr}^\0=-\frac{\d T}{T}$ and $p=4\pi T$ in comparing with \eqref{pert}. This clearly solves \eqref{explicit} with $\o=0$.

Now the crucial step is to promote temperature to be a slowly varying function of $x$: $\d T(x)=e^{i\e k_i x^i}\d T$, with $\e$ being a book-keeping parameter. The promoted function then no longer satisfies bulk equation, but needs to be corrected with $\o$ and $k$ dependence. The correction can be done order by order in $\e$. The corrected solution assumes the following form \cite{1710.04221}
\begin{align}\label{corrected}
  &h_{\m\n}=e^{-i\o v+i\e k_i x^i}\(h_{\m\n}^{RT}(x)+\e h_{[1]\m\n}(x)+\e^2 h_{[2]\m\n}(x)+\cdots\), \no
  &a_\m=e^{-i\o v+i\e k_i x^i}\(\e a_{[1]\m\n}(x)+\e^2 a_{[2]\m\n}(x)+\cdots\).
\end{align}
Written explicitly in terms of the fields in \eqref{pert} and the constant $\d T$, we have
\begin{align}\label{eps_expand}
  &h_{ij}^\0=e^{i\e k_i x^i}\(\frac{\pd g_{ij}^\0}{\pd T}\d T+h_{[1]ij}^0+\cdots\),\no
  &v_i=e^{i\e k_i x^i}\(\e v_{[1]i}+\e^2 v_{[2]i}+\cdots\), \no
  &p=e^{i\e k_i x^i}\(4\pi \d T+\e p_{[1]}+\cdots\), \no
  &\o=\e \o_{[1]}+\e^2\o_{[2]}+\cdots.
\end{align}
Note that except for $p$ and $h_{ij}$, which are nonvanishing at $O(\e^0)$ in thermodynamic normal mode \eqref{RT}, all other quantities begin with $O(\e)$. The power counting is also consistent with our MHD analysis, where $\d T/v_x\sim k/\o\sim 1/k\sim 1/\e$. It holds for $k\perp v$, which we assume below. For $k\parallel v$, a different power counting needs to be assumed.
Now we can plug \eqref{eps_expand} into \eqref{explicit} and solve it order by order in $\e$ to determine the dispersion relation $\o(k)$. In fact, we only need the first equation in \eqref{explicit}, which is from energy conservation. Furthermore, since our background is homogeneous, coefficients of the expansion in \eqref{eps_expand} are constants, thus we would have algebraic equations. Following DGZ, we have written the energy conservation equation as conservation of thermal current.
At $O(\e)$, it gives 
\begin{align}\label{omg1}
  i\o_{[1]} g^{ij}_\0\frac{\pd h_{ij}^\0}{\pd T}=0,
\end{align}
thus $\o_{[1]}=0$. At $O(\e^2)$, noting that $T(x)=T+e^{i\e k_i x^i}\d T$, we obtain
\begin{align}\label{omg2}
  2ik_i v_{[1]}^i=i\o_{[2]}g^{ij}_\0\frac{\pd h_{ij}^\0}{\pd T}.
\end{align}
To solve for $\o_{[2]}$, we need to know $v_{[1]}^i$. This can be obtained in the following way. To first order in $\e$, $v_{[1]}^i$ simply gives the horizon thermal current $Q_\0^i$, which can be expressed as a response to temperature gradient:
\begin{align}
  4\pi T\sqrt{-g_\0}\e v_{[1]}^i=Q_\0^i=-{\k}_H^{ij}\pd_j\(T(x)\)=-i\e{\k}_H^{ij}k_j\d T,
\end{align}
where ${\k}_H$ is the horizon thermal conductivity. 
Plugging this into \eqref{omg2}, we obtain the dispersion relation
\begin{align}\label{disp}
  \o_{[2]}=\frac{-ik_ik_j{\k}_H^{ij}}{c_v},
\end{align}
where we have used $c_v=T\frac{\pd s}{\pd T}=\frac{4\pi T\pd \sqrt{-g_\0}}{\pd T}$. Note that we have $k\perp v$ and also $k\perp B$. From this, we extract a diffusion constant $D_T=\frac{{\k}}{c_v}$. Here ${\k}$ corresponds to transverse thermal conductivity and it has been shown that it does not flow along the radial direction \cite{Donos:2014cya} so that we omit the subscript H.
%

Finally we show \eqref{disp} actually matches with our result from magnetohydrodynamics. First, we recall the Kubo formula that derived from MHD \cite{1703.08757}
\begin{align}
\lim \limits_{\omega \to 0}\frac{1}{\omega} Im G_{T_{tx}T_{tx}}=\frac{w^2}{\sigma _\bot B^2},
\end{align}
and the definition of thermal conductivity from thermal current
\begin{align}
\kappa =\frac{1}{T}\lim \limits_{\omega \to 0}\frac{1}{\omega} Im G_{Q_{x}Q_{x}}=\frac{1}{T}\lim \limits_{\omega \to 0}\frac{1}{\omega} Im G_{T_{tx}T_{tx}}=\frac{1}{T}\frac{w^2}{\sigma _\bot B^2}.
\end{align}
where $w=\varepsilon +p=Ts$ by the first law of thermodynamics. Therefore the thermal diffusion constant is
\begin{align}
D_T=\frac{\kappa}{c_v}=\frac{w^2}{\sigma _{\perp}B^2\frac{\partial s}{\partial T}T^2}.
\end{align}
This is in perfect agreement with our MHD result in the previous section.

We will relate the thermal diffusion constant to chaotic quantities in the next section. It is instructive to express $D_T=\k/c_v$ in terms of metric functions. Using $s=4\pi\sqrt{V_{x}^2(r_h)V_{y}(r_h)}$ and $\s_\perp=\sqrt{V_{y}(r_h)}$ \cite{Li:2018ufq}, we can express $\k$ entirely with metric in the IR fixed point.
Since $c_v=\frac{\pd s}{\pd T}$, $D_T$ can then be expressed as
\begin{align}\label{DT}
  D_T=8\pi \frac{V_{x}(r)}{B^2}\(\frac{2}{V_{x}(r)}\frac{\pd V_{x}(r)}{\pd T}+\frac{1}{V_{y}(r)}\frac{\pd V_{y}(r)}{\pd T}\)|_{r=r_h}.
\end{align}

\section{Shock Wave and Quantum Chaos}

The chaotic quantities including Lyapunov exponent and butterfly velocity can be calculated using shock wave solution in magnetic brane background \cite{1409.8180}.
Since magnetic brane is anisotropic, we adopt the method in \cite{1705.07896}. In Kruskal coordinates the anisotropic metric is given by
\begin{equation}
\mathrm{d}s^2=A(uv)\mathrm{d}u\mathrm{d}v+V_x(uv)\Big((\mathrm{d}x^1)^2+(\mathrm{d}x^2)^2\Big)+V_y(uv)\mathrm{d}y^2,
\end{equation}
where $uv=-e^{U^\prime(r_h)r_*(r)},~u/v=-e^{-U^\prime(r_h)t}$, with $r_*$ being the tortoise coordinate defined by $\mathrm{d}r_*=\mathrm{d}r/U(r)$. In addition, $A(uv)=\frac{4U(r)}{uvU^\prime(r_h)^2}$. Note that, in this coordinate the horizon is at $u=v=0,~r=r_h$.

Consider an operator insertion at time $t$ to the thermal field double state, this is dual to particle injection to the background in gravitational description. The boundary time evolution corresponds a boost of the particle energy in the Kruskal coordinates. In particular, particle with energy $E_0$ will have energy $E_0e^{\frac{2\pi}{\beta}t}$ in local time. When $e^{\frac{2\pi}{\beta}t}\sim O(N^2)$, the backreaction of particle to the background geometry can no longer be neglected. This fixes the scrambling time $t_*\sim\beta\log N^2$. The backreacted geometry is given by a shock wave localized at the horizon in the original background. 
In Kruskal coordinates the resulting stress tensor of particle is given by
\begin{equation}
(\delta T_{uu})_{particle}\sim E_0e^{\frac{2\pi}{\beta}t}\delta(u)\delta(\vec{x}).
\end{equation}
The corresponding shock wave background is given by
\begin{equation}
\mathrm{d}s^2=A(uv)\mathrm{d}u\mathrm{d}v-A(uv)\delta(u)h(\vec{x})\mathrm{d}u^2+V_x(uv)\Big((\mathrm{d}x^1)^2+(\mathrm{d}x^2)^2\Big)+V_y(uv)\mathrm{d}y^2.
\end{equation}
The perturbed Einstein equation can be written as
\begin{equation}
\delta G_{uu}=(\delta T_{uu})_{particle},
\end{equation}
which leads to the following equation of $h(\vec{x})$
\begin{equation}\label{poisson}
\Bigg(\frac{1}{V_x(r_h)}\partial_{x_1}^2+\frac{1}{V_x(r_h)}\partial_{x_2}^2+\frac{1}{V_y(r_h)}\partial_y^2-m^2\Bigg)h(\vec{x})\sim\frac{16\pi G_N}{A(0)}E_0e^{\frac{2\pi}{\beta}t}\delta(\vec{x}),
\end{equation}
with the effective mass $m^2$ given by
\begin{equation}\label{mass}
m^2=\frac{1}{A(uv)}\Bigg(\frac{2V_x^\prime(uv)}{V_x(uv)}+\frac{V_y^\prime(uv)}{V_y(uv)}\Bigg)\Bigg|_{u=0}=\pi T\Bigg(\frac{2V_x^\prime(r_h)V_y(r_h)+V_y^\prime(r_h)V_x(r_h)}{V_x(r_h)V_y(r_h)}\Bigg).
\end{equation}
\eqref{poisson} is a type of Poisson equation, which can be solved as
\begin{equation}\label{h_sol}
h(x_1,x_2,y)\sim\frac{E_0e^{\frac{2\pi}{\beta}(t-t_*)-m|x|}}{|x|},
\end{equation}
where $|x|$ is given by
\begin{equation}
|x|=\sqrt{V_x(r_h)x_1^2+V_x(r_h)x_2^2+V_y(r_h)y^2}.
\end{equation}
The exponent in \eqref{h_sol} gives both Lyapunov exponent and butterfly velocity. The Lyapunov exponent $\l_L=2\pi T$ saturating the MSS bound as expected. The butterfly velocity is anisotropic.
To calculate it in the $x_1$ direction, we can set $x_2=y=0$ to get
\begin{equation}
v_{x1}^2=\frac{4\pi^2T^2}{V_x(r_h)m^2}.
\end{equation}
Similarly, we can get in different directions
\begin{equation}\label{butterfly}
v_{x1}^2=v_{x2}^2=\frac{4\pi^2T^2}{V_x(r_h)m^2},\qquad v_{y}^2=\frac{4\pi^2T^2}{V_y(r_h)m^2}.
\end{equation}

\section{Thermal Diffusion Bound}

\begin{figure}
\includegraphics[width=0.8\textwidth]{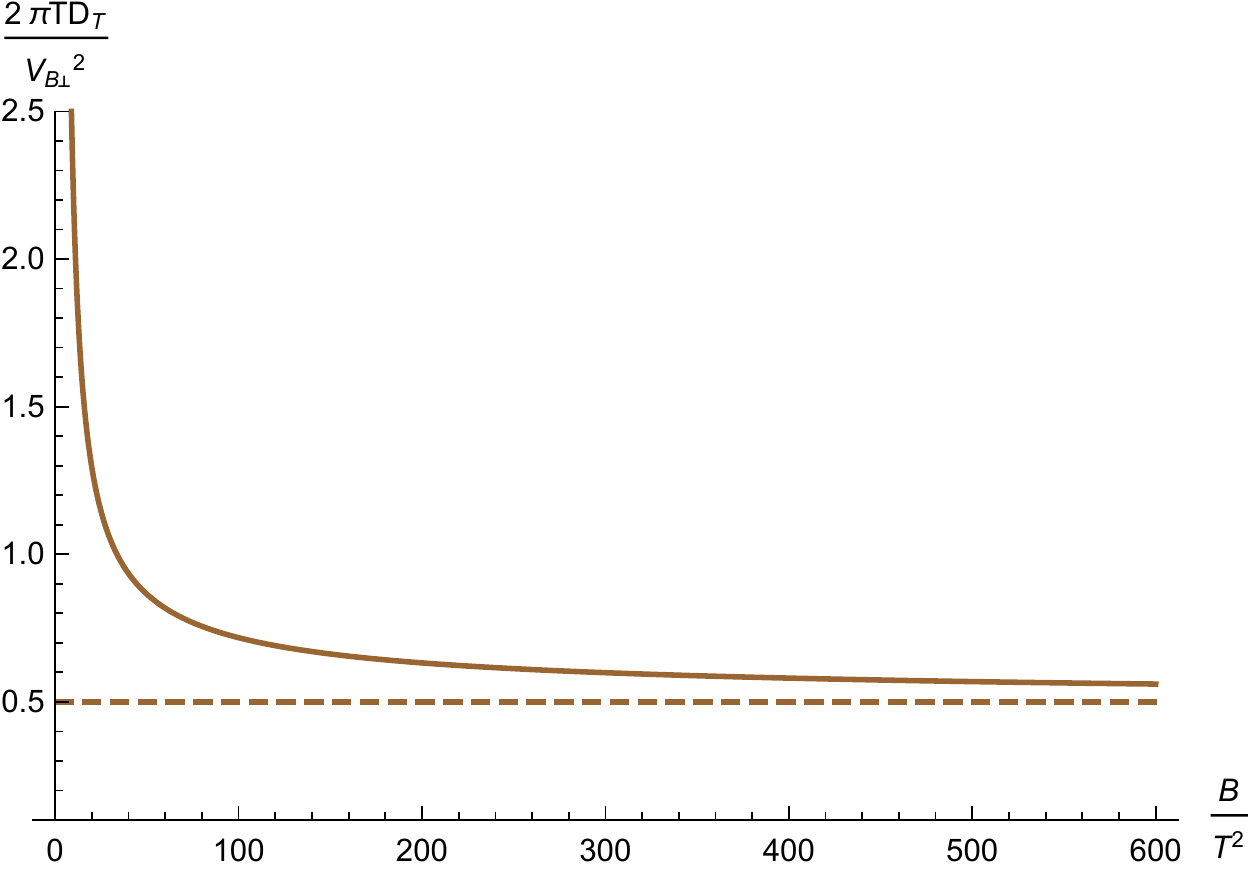}
\caption{\label{fig_bound}The ratio $\l_LD_T/v_B^2$ as a function of $B/T^2$. It is a monotonically decreasing function with an asymptotic value of $1/2$. The asymptotic value can be obtained from dimensionally reduced metric $BTZ\times R^2$ in the large $B$ limit.}
\end{figure}

Now we are ready to verify the bound on thermal diffusion constant from \eqref{DT} and \eqref{butterfly}. Note that the latter depends only on horizon metric, but the former also depends on variation of horizon metric with temperature, which needs to be obtained from numerical solution of magnetic brane background. To test the bound, we calculate the ratio $\l_LD_T/v_B^2$:
\begin{align}\label{bound_ratio}
  \l_LD_T/v_B^2=4\pi\frac{V_x(r)^2}{B^2}\(\frac{2V_x'(r)}{V_x(r)}+\frac{V_y'(r)}{V_y(r)}\)\(\frac{2\pd V_x(r)/\pd T}{V_x(r)}+\frac{\pd V_y(r)/\pd T}{V_y(r)}\)^{-1}|_{r=r_h}.
\end{align}
Here we take the transverse butterfly velocity. We show the dependence of the ratio on external magnetic field in Fig.~\ref{fig_bound}. It is a monotonically decreasing function with an asymptotic value of $1/2$.
In fact the asymptotic value can be confirmed analytically from the large $B$ limit. In this case, the background is simply given by $BTZ\times R^2$ from dimensional reduction \cite{0908.3875}. With our standard normalization for electromagnetic field, the explicit metric is given by:
\begin{align}\label{BTZ}
  ds^2=-3(r^2-r_h^2)dt^2+\frac{dr^2}{3(r^2-r_h^2)}+\frac{B}{2\sqrt{3}}\((dx^1)^2+(dx^2)^2\)+3r^2dy^2.
\end{align}
Noting that $r_h=2\pi T/3$, we can easily verify that $\l_LD_T/v_B^2=1/2$.

It is instructive to compare the case of magnetized plasma with the case of incoherent metal \cite{1604.01754}. The external magnetic field in our case exerting Lorentz force to the plasma charge carriers, introducing momentum dissipation just as lattice in metal. The large $B$ limit corresponds to the incoherent metal limit. However the asymptotic value of the ratio $\l_LD_T/v_B^2=1$ in incoherent metal case differs from ours.
In a sense, the incoherent metal case is closer to 4D magnetic black hole, which is dual to $2+1D$ plasma subject to external magnetic field. We have verified in appendix that in the 4D magnetic black hole case the ratio $\l_LD_T/v_B^2$ approaches $1$ from above. In fact the 4D magnetic black hole has the same IR fixed point $AdS_2\times R^2$ as that of incoherent metal. In contrast the 5D magnetic black hole (brane) has anisotropic IR fixed point $BTZ\times R^2$.
Interestingly, if we consider black hole in $d+1$ dimension with maximally symmetric magnetic field as in \cite{0908.3875}, the IR fixed point is $AdS_2\times R^{d-1}$ for $d$ even and $BTZ\times R^{d-2}$ for $d$ odd. We determine the corresponding asymptotic ratio to be $1$ and $1/2$ respectively in appendix. The difference between the two cases is that the background is isotropic/anisotropic for $d$ even/odd respectively.
It is also worth noting that holographic model for metal with hyperscaling violation gives the asymptotic value $\l_LD_T/v_B^2=\frac{z}{2z-2}$, with $z$ being the dynamical critical exponent \cite{1604.01754,1705.07896}. It approaches $1/2$ as $z\to \infty$.

\section{A special point and pole-skipping phenomenon}

It has been found recently that there is a special point in energy density correlator. The special point is set entirely by chaotic quantities
\begin{align}\label{sp}
  \o=i\l_L,\quad k=i\frac{\l_L}{v_B}.
\end{align}
Note that the frequency and momentum is purely imaginary. At the special point, the energy density correlator becomes undetermined, but depends on how the special point is approached \cite{1809.01169}. The indeterminacy of the correlator is related to a special property of Einstein equation in holographic systems: the $vv$ component of Einstein equation is trivially satisfied, with $v$ being the Eddington-Finkelstein coordinate. In particularly, it has been shown that generically the Einstein equation near the horizon has more than one solution \cite{1809.01169}, which leads to indeterminacy of the energy density correlator. We will show that it is also true for magnetic brane solution, in which the special point is realized in a somewhat non-trivial way.

We start by writing the magnetic brane solution in Eddington-Finkelstein coordinate
\begin{align}\label{EF}
  ds^2=-U(r)dv^2+2dvdr+V_x(r)\((dx^1)^2+(dx^2)^2\)+V_y(r)dy^2.
\end{align}
Near the horizon, the metric functions have the following expansion
\begin{align}\label{G_exp}
  &U=4\pi T(r-r_h)+u_2(r-r_h)^2+\cdots,\\
  &V_x=1+vx_1(r-r_h)+\cdots,\no
  &V_y=1+vy_1(r-r_h)+\cdots.
\end{align}
Note that the leading terms in the expansion of $U$ and $V_x(V_y)$ are $4\pi T$ and $1$ respectively. The former is determined by the magnetic brane temperature. The latter is arbitrary. It can be realized by rescaling of $x_1$, $x_2$ and $y$ coordinates \cite{0908.3875}. The rescaling also changes the value of magnetic field, which we take as $b$. It is easy to obtain the higher order coefficients of the expansion as
\begin{align}\label{rtc}
  u_2=\frac{5b^2-24}{12},\quad vx_1=\frac{-b^2+12}{6\pi T},\quad vy_1=\frac{b^2+24}{12\pi T}.
\end{align}
Now we turn on the following fluctuations $h_{vv}(v,x_1,r)$, $h_{vr}(v,x_1,r)$, $h_{rr}(v,x_1,r)$, $h_{vx1}(v,x_1,r)$ and $h_{x1x1}(v,x_1,r)$. The fluctuations assume the following expansion near the horizon.
\begin{align}\label{h_exp}
  h_{vv}=e^{i\o v+i k x_1}\(h_{vv0}+h_{vv1}(r-r_h)+\cdots\),
\end{align}
and similarly for other components.
The $vv$ component of Einstein equation can be expanded near horizon using \eqref{G_exp} and \eqref{h_exp} with the leading coefficient
\begin{align}\label{vv}
  \(\frac{k^2}{2}+\frac{i\o(b^2-24)}{16\pi T}\)h_{vv0}+\frac{2\o^2-4i\pi T\o}{4}h_{xx0}+\frac{4k\o-8i\pi Tk}{4}h_{vx0}=0.
\end{align}
We find that there is a special point at
\begin{align}\label{sp_scale}
\o=2i\pi T,\quad k=i\frac{\sqrt{24-b^2}}{2}.
\end{align}
On the other hand, we can express the butterfly velocity in terms of the metric \eqref{G_exp}. We can simply apply \eqref{butterfly} and use \eqref{rtc} to obtain
\begin{align}\label{vb_scale}
  v_{x}=\frac{4\pi T}{\sqrt{b^2-24}}.
\end{align}
We can readily verify that the special point \eqref{sp} by noting $\l_L=2\pi T$.
Note that both \eqref{sp_scale} and \eqref{vb_scale} are subject to rescaling to coordinates with proper normalization at the boundary. Nevertheless, the rescaling of $x_1$ coordinate acts on $k$ and $v_x$ in the opposite way leaving the form of \eqref{sp} invariant. The rescaling in other coordinates does not lead to any change. Therefore we confirm the existence of special point \eqref{sp} in magnetic brane background as well, independent of the strength of magnetic field.

\begin{figure}
\includegraphics[width=0.8\textwidth]{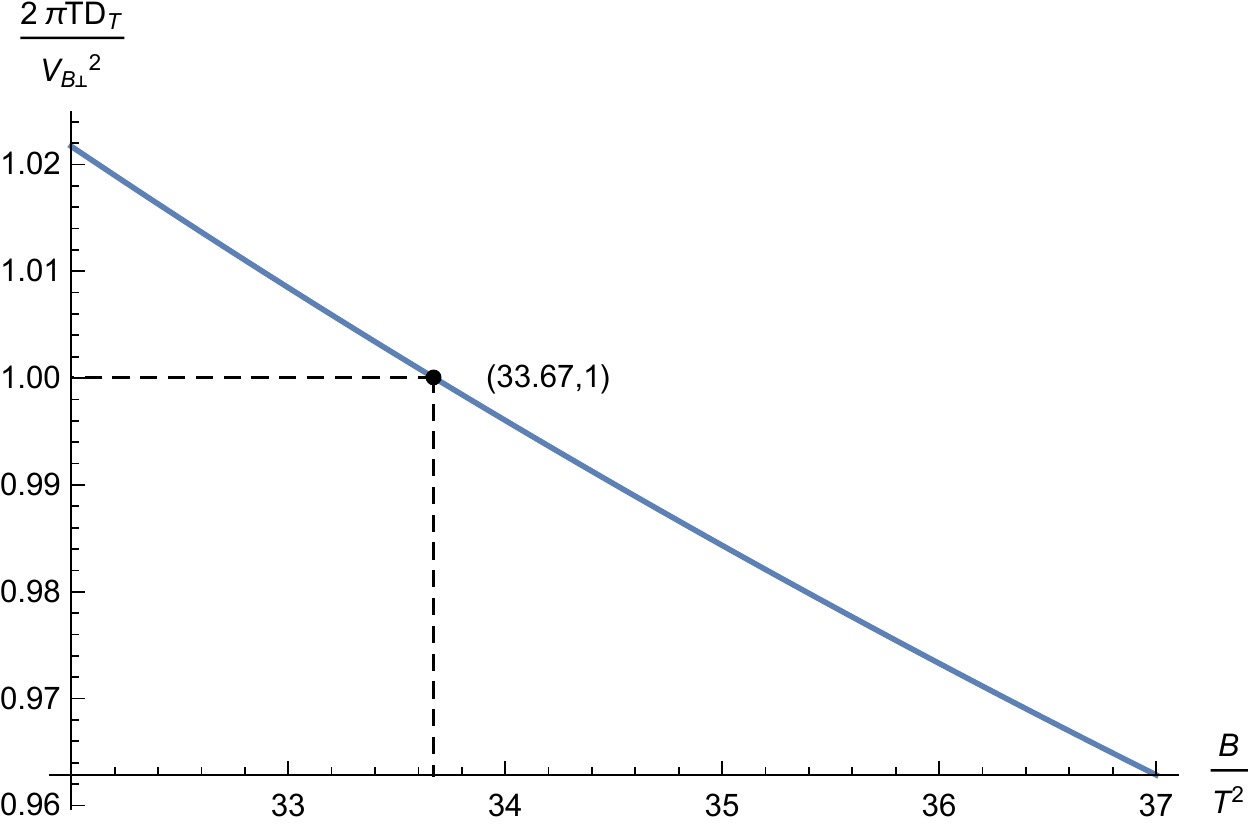}
\caption{\label{fig_deq1}The ratio $\l_LD_T/v_B^2$ approaches 1 when $B/T^2\approx33.67$. We expect near this region hydrodynamics is a good approximation.}
\end{figure}

Remarkably, the special point is found to lie on the branch of quasi-normal mode whose low momenta limit corresponds to the hydrodynamic diffusive mode \cite{1809.01169}. Note that for purely imaginary momenta, the diffusive mode becomes unstable. In particular, the hydrodynamic description of the unstable mode seems more and more accurate in the incoherent metal limit \cite{1809.01169} in the following sense: the ratio $D\l_L/v_B^2$ approaches one. In other words, the chaotic diffusion constant defined by the special point tends to the hydrodynamic diffusion constant at low energy:
\begin{align}
  D=\frac{i\l_L}{i\(i\l_L/v_B\)^2}\to \frac{\o}{ik^2}|_{k\to0}.
\end{align}
In our magnetized plasma case, we find a little different result. The ratio $D\l_L/v_x^2$ approaches one half in the limit of large magnetic field. This limit is analogous to incoherent metal limit as both correspond to large momentum dissipation. It seems that in our case the chaotic diffusion constant and hydrodynamic diffusion constant agrees at a finite value of magnetic field. Our numerical result in Fig.~\ref{fig_deq1} indicates that it occurs at $B/T^2=33.67$. It would be interesting to explore properties of MHD near this value of magnetic field. However this interpretation comes with certain caveats. We have not confirmed that the special point lie indeed on the branch of quasi-normal mode whose low momenta limit corresponds to the hydrodynamic diffusive mode. Also, we have not considered high order corrections to the diffusive mode. We leave them for future studies.

\section{Summary and outlook}

We calculate the transverse thermal diffusion constant $D_T$  in magnetic brane background. We also calculate the Lyapunov exponent $\l_L$ and butterfly velocity $v_B$ in the same background. We find Blake's bound on thermal diffusion constant holds in magnetic brane background as well. In particular, the ratio $D_T\l_L/v_{B\perp}^2$ is a monotocially decreasing function of $B/T^2$ with an asymptotic value of $1/2$.

We also find the existence of special point $\o=i\l_L$ and $k=i\l_L/v_x$. At this point, the energy density correlator becomes undetermined, similar to the case of incoherent metal. There is one interesting difference from the incoherent metal case: in the limit of strong momentum dissipation, the ratio $D_T\l_L/v_{B\perp}^2$ tends to $1/2$ in our case, while it approaches $1$ in incoherent metal case. The deviation from $1$ in our case seems due to anisotropy of the background. Similar deviation of the asymptotic value of the ratio from one is also seen in holographic models with hyperscaling violation. It would be interesting to explore further magnetohydrodynamics at large $B$ and shock wave in the $BTZ\times R^2$ background where analytic calculations are possible.

Last but not the least, the present analysis can be readily generalized to the charged magnetic brane background dual to charged magnetized plasma, in which mixing between charge diffusion and thermal diffusion is expected \cite{1405.3651}. It would offer further non-trivial test of the intriguing connection between low energy transport and chaos.

\begin{acknowledgments}
It is a pleasure to thank Xian Gao, Xian-hui Ge, Yan Liu, Jia-rui Sun and especially Wei-jia Li for useful discussions. S.L. would like to thank University of Science and Technology of China for hospitality at the final stage of this work. S.L. is supported by One Thousand Talent Program for Young Scholars and NSFC under Grant Nos 11675274 and 11735007. J.J.M would like to thank Sun Yat-Sen University for hospitality during which part of this work has been completed.

\end{acknowledgments}

\appendix

\section{Diffusion bound in $2+1D$ and higher dimensions}
Here we consider the energy diffusion in $2+1 D$ plasma, which is dual to $AdS_4$ black hole with external magnetic field.
\begin{equation}
S=\frac{1}{16\pi G_4}\int\mathrm{d}^4x\sqrt{-g}\left[R+\frac{6}{L^2}-\frac{1}{4}F^{\mu\nu}F_{\mu\nu}\right].
\end{equation}
The neutral magnetized plasma solution is given by \cite{0908.3875}
\begin{align}
&\mathrm{d}s^2=-U(r)\mathrm{d}t^2+\frac{\mathrm{d}r^2}{U(r)}+r^2\mathrm{d}x^2+r^2\mathrm{d}y^2\notag\\
&A=Bx\mathrm{d}y,\quad U(r)=r^2+\frac{B^2}{4r^2}+\frac{1}{r}\left(\frac{-B^2-4r_h^4}{4r_h}\right).
\end{align}
\begin{figure}
\includegraphics[width=0.8\textwidth]{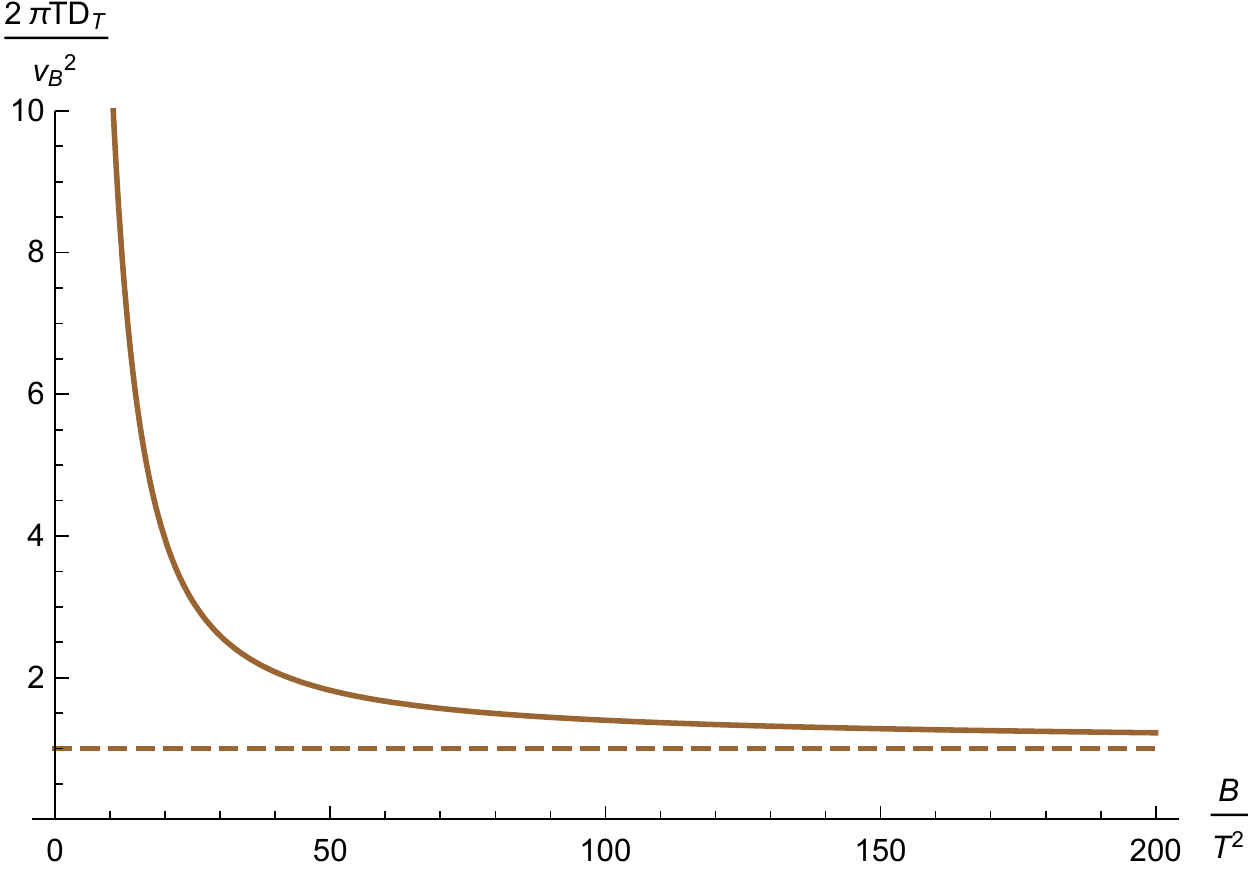}
\caption{\label{fig_bound2}The ratio $\l_LD_T/v_B^2$ as a function of $B/T^2$. The dashed line represents constant $1$. The bound is a monotonically decreasing function with an asymptotic value of $1$.}
\end{figure}
Here we set $16\pi G_4=1,L=1$. So the thermal conductivity and the specific heat directly read \cite{1705.07896}
\begin{equation}
{\kappa}=\frac{4\pi sTr_h^2}{B^2},\quad c_v=T\left(\frac{\partial s}{\partial T}\right)_B=8\pi Tr_h\left(\frac{\partial r_h}{\partial T}\right)_B.
\end{equation} 
Using the definition of black hole temperature we get the horizon radius as a root of the following equation
\begin{equation}
-\frac{B^2+2r_h^2(-6r_h^2+8\pi r_hT)}{16\pi r_h^3}=0,
\end{equation}
from which we further get
\begin{equation}
\left(\frac{\partial r_h}{\partial T}\right)_B=\frac{16\pi r_h}{12r_h+3B^2}.
\end{equation}
So thermal diffusion constant $D_T$ read
\begin{equation}
  D_T=\frac{3}{8 r_h} \left(\frac{4 r_h^4}{B^2}+1\right).
\end{equation}
Butterfly velocity in this case is $v_B^2=\pi T/r_h^2$, so the ratio of the diffusion bound is given by
\begin{equation}
\frac{2\pi TD_T}{v_B^2}=\frac{3 r_h^4}{B^2}+\frac{3}{4}.
\end{equation}
When $B/T^2\gg1$, $r_h$ approaches a constant $r_h^2\to r_0^2=\frac{B}{\sqrt{12}}$, near horizon geometry will become $AdS_2\times R^2$
\begin{equation}
\mathrm{d}s^2=-6(r-r_0)^2\mathrm{d}t^2+\frac{\mathrm{d}r^2}{6(r-r_0)^2}+\frac{B}{\sqrt{12}}\mathrm{d}x^2+\frac{B}{\sqrt{12}}\mathrm{d}y^2.
\end{equation}
So the bound ratio is $2\pi TD_T/v_B^2=1$. When $B/T^2$ is finite, we present the numerical result of the bound in Fig.~\ref{fig_bound2}, the asymptotic value matches the analysis of near horizon geometry $AdS_2\times R^2$ in the large $B$ limit.

For higher dimensional magnetic brane, we consider only limiting case of large and maximally symmetric magnetic field \cite{0908.3875}. When $d$ is even, it has a IR fixed point solution, which is just $BTZ\times R^{d-2}$
\begin{equation}
ds^2=-(d-1)(r^2-r_h^2)dt^2+\frac{1}{(d-1)(r^2-r_h^2)}dr^2+(d-1)r^2dy^2+\frac{B}{2\sqrt{d-1}}((dx^1)^2+\cdots (dx^{d-2})^2),
\end{equation}
with
\begin{align}
  F_{12}=F_{34}=\cdots=B\to\infty, \quad F_{iy}=0.
\end{align}
The entropy is given by
\begin{align}
s&=4\pi \sqrt {g_{x_1x_1}(r_h)\cdots g_{x_{d-2}x_{d-2}}(r_h)g_{yy}(r_h)}\\
&=8\pi ^2T \(\frac{B}{2}\)^{\frac{d}{2}-1}(d-1)^{-\frac{d}{4}}
\end{align}
Transverse conductivity is
\begin{align}
\sigma _{\perp}&=\sqrt {g_{x_1x_1}(r_h)\cdots g_{x_{d-2}x_{d-2}}(r_h)g_{yy}(r_h)}g^{x_1x_1}(r_h)\\
&=2\pi T\(\frac{B}{2}\)^{\frac{d}{2}-2}(d-1)^{\frac{2-d}{4}}
\end{align}
So thermal diffusion constant is
\begin{align}
D_T&=\frac{w^2}{\sigma _{\perp}B^2\frac{\partial s}{\partial T}T^2}\\
&=\frac{2\pi T}{B\sqrt{d-1}}
\end{align}
It is easy to verify that the expression of butterfly velocity remain the same
\begin{align}
v_{x_1}^2&=\frac{4\pi ^2 T^2}{g_{xx}(r_h)m^2}\\
&=\frac{4\pi Tg_{yy}(r_h)}{g_{x_1x_1}(r_h)g'_{yy}(r_h)}\\
&=\frac{8\pi ^2T^2}{B\sqrt{d-1}}
\end{align}
Then the ratio in this case is
\begin{align}
\lambda _L D_T /v_B ^2=\frac{1}{2}
\end{align}

When $d$ is odd, it is an isotropic solution. And when $B/T^2\to\infty$, horizon radius $r_h^2\to r_+^2=B/\sqrt{d}$. In this case the near horizon geometry is just $AdS_{2}\times R^{d-1}$. It has been proved in \cite{1611.09380} that when the near horizon geometry is $AdS_{2}\times R^{d-1}$, the ratio of thermal diffusion $\lambda _L D_T /v_B ^2=1$.


\begin{thebibliography}{90}

\bibitem{1405.3651} 
  S.~A.~Hartnoll,
  Nature Phys.\  {\bf 11}, 54 (2015)
  doi:10.1038/nphys3174
  [arXiv:1405.3651 [cond-mat.str-el]].



\bibitem{1306.0622} 
  S.~H.~Shenker and D.~Stanford,
  JHEP {\bf 1403}, 067 (2014)
  doi:10.1007/JHEP03(2014)067
  [arXiv:1306.0622 [hep-th]].



\bibitem{1409.8180} 
  D.~A.~Roberts, D.~Stanford and L.~Susskind,
  JHEP {\bf 1503}, 051 (2015)
  doi:10.1007/JHEP03(2015)051
  [arXiv:1409.8180 [hep-th]].



\bibitem{1412.6087} 
  S.~H.~Shenker and D.~Stanford,
  JHEP {\bf 1505}, 132 (2015)
  doi:10.1007/JHEP05(2015)132
  [arXiv:1412.6087 [hep-th]].



\bibitem{1503.01409} 
  J.~Maldacena, S.~H.~Shenker and D.~Stanford,
  JHEP {\bf 1608}, 106 (2016)
  doi:10.1007/JHEP08(2016)106
  [arXiv:1503.01409 [hep-th]].



\bibitem{1605.06098} 
  K.~Jensen,
  Phys.\ Rev.\ Lett.\  {\bf 117}, no. 11, 111601 (2016)
  doi:10.1103/PhysRevLett.117.111601
  [arXiv:1605.06098 [hep-th]].



\bibitem{1603.08510} 
  M.~Blake,
  Phys.\ Rev.\ Lett.\  {\bf 117}, no. 9, 091601 (2016)
  doi:10.1103/PhysRevLett.117.091601
  [arXiv:1603.08510 [hep-th]].



\bibitem{1604.01754} 
  M.~Blake,
  Phys.\ Rev.\ D {\bf 94}, no. 8, 086014 (2016)
  doi:10.1103/PhysRevD.94.086014
  [arXiv:1604.01754 [hep-th]].



\bibitem{1610.02669} 
  Y.~Ling, P.~Liu and J.~P.~Wu,
  JHEP {\bf 1710}, 025 (2017)
  doi:10.1007/JHEP10(2017)025
  [arXiv:1610.02669 [hep-th]].



\bibitem{1611.09380} 
  M.~Blake and A.~Donos,
  JHEP {\bf 1702}, 013 (2017)
  doi:10.1007/JHEP02(2017)013
  [arXiv:1611.09380 [hep-th]].



\bibitem{1702.08803} 
  S.~F.~Wu, B.~Wang, X.~H.~Ge and Y.~Tian,
  Phys.\ Rev.\ D {\bf 97}, no. 10, 106018 (2018)
  doi:10.1103/PhysRevD.97.106018
  [arXiv:1702.08803 [hep-th]].



\bibitem{1704.00947} 
  K.~Y.~Kim and C.~Niu,
  JHEP {\bf 1706}, 030 (2017)
  doi:10.1007/JHEP06(2017)030
  [arXiv:1704.00947 [hep-th]].



\bibitem{1705.07896} 
  M.~Blake, R.~A.~Davison and S.~Sachdev,
  Phys.\ Rev.\ D {\bf 96}, no. 10, 106008 (2017)
  doi:10.1103/PhysRevD.96.106008
  [arXiv:1705.07896 [hep-th]].



\bibitem{1705.01766} 
  M.~Baggioli and W.~J.~Li,
  JHEP {\bf 1707}, 055 (2017)
  doi:10.1007/JHEP07(2017)055
  [arXiv:1705.01766 [hep-th]].



\bibitem{1707.02843} 
  Y.~Ling and Z.~Y.~Xian,
  JHEP {\bf 1709}, 003 (2017)
  doi:10.1007/JHEP09(2017)003
  [arXiv:1707.02843 [hep-th]].



\bibitem{1708.05691} 
  D.~Giataganas, U.~Gürsoy and J.~F.~Pedraza,
  Phys.\ Rev.\ Lett.\  {\bf 121}, no. 12, 121601 (2018)
  doi:10.1103/PhysRevLett.121.121601
  [arXiv:1708.05691 [hep-th]].



\bibitem{1708.08822} 
  H.~S.~Jeong, Y.~Ahn, D.~Ahn, C.~Niu, W.~J.~Li and K.~Y.~Kim,
  JHEP {\bf 1801}, 140 (2018)
  doi:10.1007/JHEP01(2018)140
  [arXiv:1708.08822 [hep-th]].



\bibitem{1712.00705} 
  X.~H.~Ge, S.~J.~Sin, Y.~Tian, S.~F.~Wu and S.~Y.~Wu,
  JHEP {\bf 1801}, 068 (2018)
  doi:10.1007/JHEP01(2018)068
  [arXiv:1712.00705 [hep-th]].



\bibitem{1805.05351} 
  D.~Avila, V.~Jahnke and L.~Patiño,
  JHEP {\bf 1809}, 131 (2018)
  doi:10.1007/JHEP09(2018)131
  [arXiv:1805.05351 [hep-th]].



\bibitem{1609.07832} 
  Y.~Gu, X.~L.~Qi and D.~Stanford,
  JHEP {\bf 1705}, 125 (2017)
  doi:10.1007/JHEP05(2017)125
  [arXiv:1609.07832 [hep-th]].



\bibitem{1702.08462} 
  Y.~Gu, A.~Lucas and X.~L.~Qi,
  SciPost Phys.\  {\bf 2}, no. 3, 018 (2017)
  doi:10.21468/SciPostPhys.2.3.018
  [arXiv:1702.08462 [hep-th]].



\bibitem{1705.09818} 
  Y.~Chen, H.~Zhai and P.~Zhang,
  JHEP {\bf 1707}, 150 (2017)
  doi:10.1007/JHEP07(2017)150
  [arXiv:1705.09818 [hep-th]].



\bibitem{1711.07903} 
  W.~Cai, X.~H.~Ge and G.~H.~Yang,
  JHEP {\bf 1801}, 076 (2018)
  doi:10.1007/JHEP01(2018)076
  [arXiv:1711.07903 [hep-th]].



\bibitem{1904.02174} 
  H.~Guo, Y.~Gu and S.~Sachdev,
  arXiv:1904.02174 [cond-mat.str-el].



\bibitem{1612.05500} 
  M.~Baggioli, B.~Goutéraux, E.~Kiritsis and W.~J.~Li,
  JHEP {\bf 1703}, 170 (2017)
  doi:10.1007/JHEP03(2017)170
  [arXiv:1612.05500 [hep-th]].



\bibitem{1611.00003} 
  A.~A.~Patel and S.~Sachdev,
  Proc.\ Nat.\ Acad.\ Sci.\  {\bf 114}, 1844 (2017)
  doi:10.1073/pnas.1618185114
  [arXiv:1611.00003 [cond-mat.str-el]].



\bibitem{1612.00849} 
  R.~A.~Davison, W.~Fu, A.~Georges, Y.~Gu, K.~Jensen and S.~Sachdev,
  Phys.\ Rev.\ B {\bf 95}, no. 15, 155131 (2017)
  doi:10.1103/PhysRevB.95.155131
  [arXiv:1612.00849 [cond-mat.str-el]].



\bibitem{1608.03286} 
  A.~Lucas and J.~Steinberg,
  JHEP {\bf 1610}, 143 (2016)
  doi:10.1007/JHEP10(2016)143
  [arXiv:1608.03286 [hep-th]].



\bibitem{1710.07896} 
  W.~J.~Li, P.~Liu and J.~P.~Wu,
  JHEP {\bf 1804}, 115 (2018)
  doi:10.1007/JHEP04(2018)115
  [arXiv:1710.07896 [hep-th]].



\bibitem{1710.03738} 
  A.~Mokhtari, S.~A.~Hosseini Mansoori and K.~Bitaghsir Fadafan,
  Phys.\ Lett.\ B {\bf 785}, 591 (2018)
  doi:10.1016/j.physletb.2018.09.020
  [arXiv:1710.03738 [hep-th]].



\bibitem{0908.3875} 
  E.~D'Hoker and P.~Kraus,
  JHEP {\bf 0910}, 088 (2009)
  doi:10.1088/1126-6708/2009/10/088
  [arXiv:0908.3875 [hep-th]].



\bibitem{1710.00921} 
  S.~Grozdanov, K.~Schalm and V.~Scopelliti,
  Phys.\ Rev.\ Lett.\  {\bf 120}, no. 23, 231601 (2018)
  doi:10.1103/PhysRevLett.120.231601
  [arXiv:1710.00921 [hep-th]].



\bibitem{1801.00010} 
  M.~Blake, H.~Lee and H.~Liu,
  JHEP {\bf 1810}, 127 (2018)
  doi:10.1007/JHEP10(2018)127
  [arXiv:1801.00010 [hep-th]].



\bibitem{1809.01169} 
  M.~Blake, R.~A.~Davison, S.~Grozdanov and H.~Liu,
  JHEP {\bf 1810}, 035 (2018)
  doi:10.1007/JHEP10(2018)035
  [arXiv:1809.01169 [hep-th]].



\bibitem{1904.12883} 
  M.~Blake, R.~A.~Davison and D.~Vegh,
  arXiv:1904.12883 [hep-th].



\bibitem{1904.12862} 
  S.~Grozdanov, P.~K.~Kovtun, A.~O.~Starinets and P.~Tadić,
  arXiv:1904.12862 [hep-th].



\bibitem{1703.08757} 
  J.~Hernandez and P.~Kovtun,
  JHEP {\bf 1705}, 001 (2017)
  doi:10.1007/JHEP05(2017)001
  [arXiv:1703.08757 [hep-th]].



\bibitem{1610.07392} 
  S.~Grozdanov, D.~M.~Hofman and N.~Iqbal,
  Phys.\ Rev.\ D {\bf 95}, no. 9, 096003 (2017)
  doi:10.1103/PhysRevD.95.096003
  [arXiv:1610.07392 [hep-th]].



\bibitem{1106.4826} 
  I.~Papadimitriou,
  JHEP {\bf 1108}, 119 (2011)
  doi:10.1007/JHEP08(2011)119
  [arXiv:1106.4826 [hep-th]].



\bibitem{1710.04221} 
  A.~Donos, J.~P.~Gauntlett and V.~Ziogas,
  JHEP {\bf 1803}, 056 (2018)
  doi:10.1007/JHEP03(2018)056
  [arXiv:1710.04221 [hep-th]].



\bibitem{Donos:2014cya} 
  A.~Donos and J.~P.~Gauntlett,
  JHEP {\bf 1411}, 081 (2014)
  doi:10.1007/JHEP11(2014)081
  [arXiv:1406.4742 [hep-th]].



\bibitem{Li:2018ufq} 
  W.~Li, S.~Lin and J.~Mei,
  Phys.\ Rev.\ D {\bf 98}, no. 11, 114014 (2018)
  doi:10.1103/PhysRevD.98.114014
  [arXiv:1809.02178 [hep-th]].

\bibitem{1611.09380} 
  M.~Blake and A.~Donos,
  JHEP {\bf 1702}, 013 (2017)
  doi:10.1007/JHEP02(2017)013
  [arXiv:1611.09380 [hep-th]].
  
\end{thebibliography}
\end{document}